\documentclass[%
reprint,
amsmath,amssymb,
aps,
prb,
]{revtex4-1}
\usepackage{graphicx}
\usepackage{dcolumn}
\usepackage{bm}
\usepackage[usenames,dvipsnames,svgnames,table]{xcolor}
\begin{document}


\title{Liquid dynamics in partially crystalline glycerol}


\author{Alejandro Sanz}
\email[]{asanz@ruc.dk}
\author{Kristine Niss}
\affiliation{IMFUFA, Department of Science and Environment, Roskilde University, Postbox 260, DK-4000 Roskilde, Denmark}


\date{\today}

\begin{abstract}
	We present a dielectric study on the dynamics of supercooled
	glycerol during crystallization. We explore the transformation into a
	solid phase in real time by monitoring the temporal evolution of the
	amplitude of the dielectric signal. Neither the initial nucleation
	or the crystal growth influence the liquid dynamics visibly. For one
	of the samples studied, a tiny fraction of glycerol remained in the
	disordered state after the end of the transition. We examined the
	nature of the $\alpha$ relaxation in this frustrated crystal and
	find that it is virtually identical to the bulk dynamics. In
	addition, we have found no evidence that supercooled
	glycerol transforms into a peculiar phase in which either a new solid
	amorphous state or nano-crystals dispersed in a liquid matrix are
	formed.
	
\end{abstract}


\maketitle

\section{Introduction}

One of the longstanding questions in the field of condensed matter
physics is what governs the liquid-to-glass transition and the
concomitant dramatic slowing down of the transport quantities when
supercooled liquids approach the glassy state on cooling \cite{Debenedetti,Berthier,Berthier2}.
At the glass transition temperature \textit{$T_{g}$}, the liquid falls
out of thermodynamic equilibrium because the molecular rearrangements
become so slow that the equilibrium volume and enthalpy
\cite{Ediger,Dyre} cannot be reached on experimental time scales. 
In fact, the main focus in the field is the molecular relaxation taking
place in the supercooled liquid just aboe $T_g$. 
The main relaxation process, referred to as the $\alpha$ relaxation,
is cooperative in nature and  
can be measured as a response to a wide range of external
perturbations, for instance mechanical or electrical \cite{Bo}. The glass-formers community has widely
exploited the imaginary part of the complex dielectric permittivity
because of its high sensitivity to external factors such as
temperature and pressure, and also to microscopic properties of the
substance such as its degree of order, physical interactions with the
surroundings, and geometrical confinement among others
\cite{Sillren,Paluch,Aurora,Murata,Lunkenheimer2,Martin,Arndt}.

Glycerol (propane-1,2,3-triol) is one of the most studied molecular
glass-formers, principally for its extremely low tendency to
crystallize
\cite{Wuttke,Lunkenheimer3,Lunkenheimer,Schröter,Zondervan,Hayashi,Lisa}. Since glycerol normally forms a highly
stable supercooled liquid (large viscosity upon melting
\cite{Schröter}), little attention has been paid to its crystalline
phase. Even so, one can find in the literature several works mainly
focused on the high frequency dynamics of the crystalline state
\cite{Bermejo,Talon,Buchenau}. Nevertheless, real time investigations
of its transformation into solid-like structures have recently
attracted the interest of some authors \cite{Mobius,Yuan}. Inspired by
previous works on the existence of long-lived dynamic heterogeneities
in supercooled glycerol, M\"obious \textit{et al.} performed real time aging
experiments above \textit{$T_{g}$} and detected the formation of a
solid phase that showed a distinct mechanical response as compared
to the orthorhombic crystal \cite{Mobius}. This solid-like
phase presented a value of the shear modulus two orders of magnitude lower
than that of standard crystalline glycerol. In this way, M\"obious and
co-authors \textit{speculated} on the existence of a glacial phase
\cite{Ha}, that is, either a second amorphous state or a frustrated
crystal with a high degree of defects and disorder. Two
years later, by employing time-resolved neutron scattering, Yuan and
co-authors aimed to unravel the structural nature of this solid-like
phase \cite{Yuan}. They revealed, in agreement with one of the
interpretations given by M\"obious et al., the formation of a
crystalline lattice.

It is well established that there is an interplay between structure
and dynamics on the phenomenon of crystallization
\cite{Descamps,karolina,Tripathi,Sanz,Dobbertin}, though the exact nature of this is not fully
understood. The focus in this paper is on how the nucleation and growth steps of crystallization of glycerol affect the dynamics of the remaining
liquid. Moreover we study the dynamics of a partially disordered
sample, inquiring about the putative existence of a glacial phase. 

\section{Sample preparation and experimental tools}

Anhydrous glycerol ($\geq$ 99.5 \% purity,
Sigma-Aldrich$^{\circledR}$) was used without additional purification
and was manipulated, as much as we could, under nitrogen to avoid the
uptake of ambient water. In order to carry out the present study, we
utilized dielectric spectroscopy, using a sample cell
especially designed for this kind of measurement. It consists of two
parallel metal plates separated by a Kapton$^{\circledR}$ spacer of
0.25 mm thickness. With the purpose of insulating the sample cell
electrically and also from possible water contamination, the capacitor
was sealed by means of a cylindrical container made of polyether ether
ketone (\textit{PEEK}).

We filled the transducer with dry glycerol under nitrogen flow inside
a glove bag. Once the cell was properly closed, it was transferred to
the measuring cryostat. For a detailed description of the dielectric
spectroscopy set-up and sample environment control we refer the reader
to the following publication \cite{Igarashi,Igarashi2}.

\begin{figure}
	\centering
	\includegraphics[trim = 10mm 0mm 0mm 10mm, clip,width=1.0\linewidth]{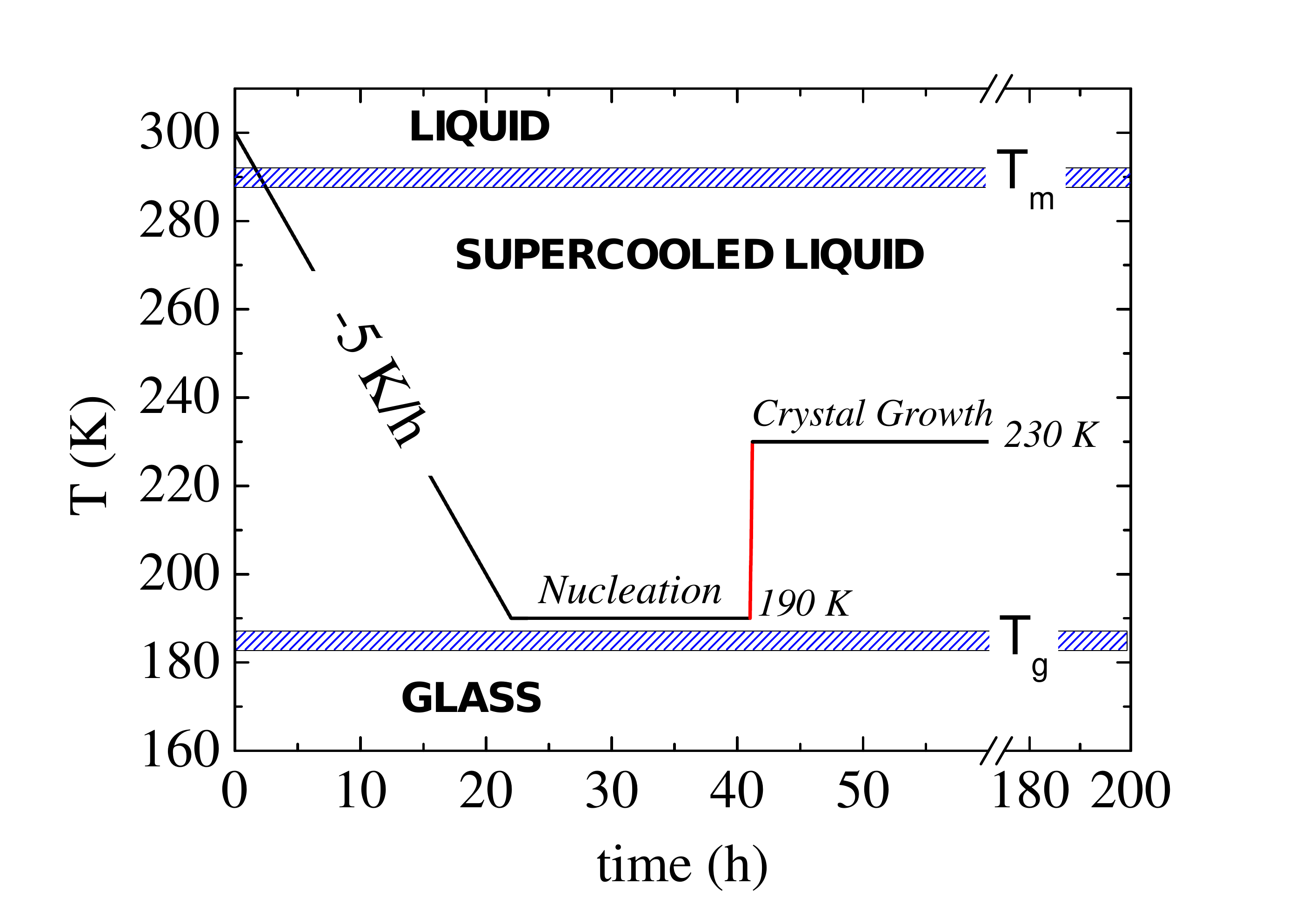}
	\caption{Thermal protocol for inducing the transformation of liquid glycerol into the crystalline phase. Nucleation of crystalline seeds is promoted by annealing at 190 K during 19 hours. Crystal growth takes place at 230 K after fast heating from 190 K.}
	\label{fig:Fig1}      
\end{figure}

The standard procedure we followed to induce the crystallization of
supercooled glycerol consisted of the thermal protocol described in
Fig. \ref{fig:Fig1}. We cooled down the sample from 300 K to 190 K at
a cooling rate of approximately 5 K/h. Then, an isothermal annealing
at 190 K was carried out. Finally, we heated up to 230 K where we
monitored isothermally the dielectric signal as crystallization occurred.

\section{Results}

\subsection{Dynamics of the fully disordered liquid}

First, we examined the dynamics of the fully disordered liquid between 195 and 240 K on heating. We explored the $\alpha$
relaxation by measuring the frequency
evolution of the complex dielectric permittivity,
$\varepsilon^*(\omega)=\varepsilon'(\omega)-i\varepsilon''(\omega)$ ,
where $\omega$ is the angular frequency, and $\varepsilon'$ and
$\varepsilon''$ are the real and imaginary parts respectively, which were in agreement with the literature \cite{Lunkenheimer,Hayashi}. We
described the relaxation curves with a dielectric version of the
$\alpha$ circuit model, also known as the dielectric version of the
Extended Bell (EB) model \cite{Saglanmak}. The complex dielectric
permittivity according to the $\alpha$ circuit model reads as follows:

\begin{equation}
	\varepsilon^{*}(\omega)=\varepsilon_{\infty}+\frac{\Delta\varepsilon}{1+\frac{1}{(i\omega\tau_{\alpha})^{-1}+k_{\alpha}(i\omega\tau_{\alpha})^{-\alpha}}}
	\label{eq:Eq1} 
\end{equation}  
Here, $\varepsilon_{\infty}$ is the instantaneous or unrelaxed dielectric constant, $\Delta\varepsilon$ is the dielectric strength, $\tau_{\alpha}$ is the $\alpha$-relaxation time, $\alpha$ is the high-frequency power law \cite{alpha} and $k_{\alpha}$ determines the broadening of the $\alpha$ peak.

\subsection{Dynamics during nucleation}

\begin{figure}
	\centering
	\includegraphics[trim = 20mm 80mm 83mm 72mm, clip, width=0.7\linewidth]{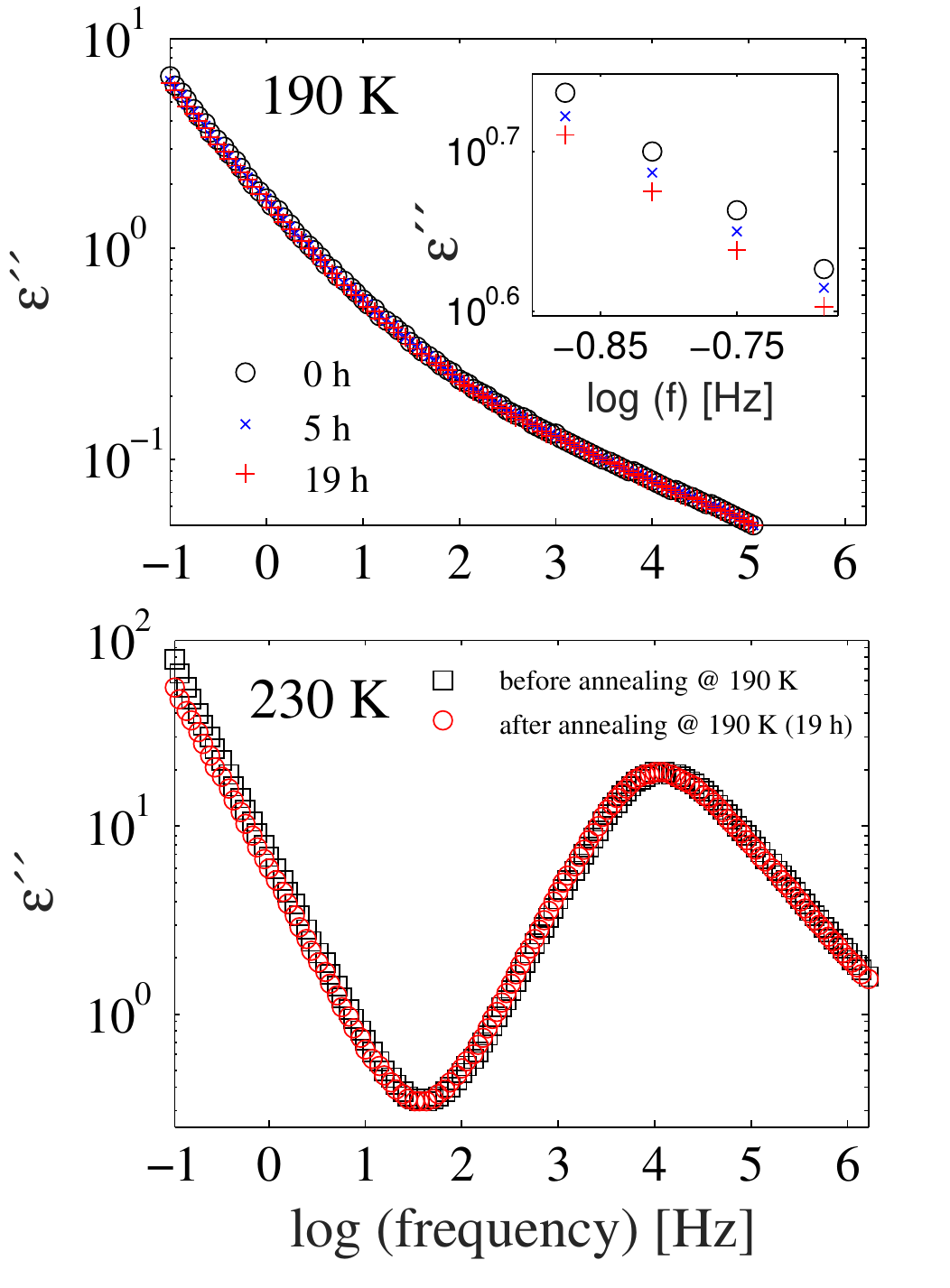}
	\caption{(Top) Frequency dependence of the imaginary part of the complex dielectric permittivity in glycerol during isothermal annealing at 190 K. The inset zooms the low frequency flank of the dielectric loss for a more detailed examination. (Bottom) Dielectric spectra of glycerol collected at 230 K, first by cooling from room temperature and second after annealing at 190 K and subsequent heating.}
	\label{fig:Fig3}
\end{figure}

The main aim of this work is to explore the
dynamics-structure relationships during the transformation of liquid
glycerol into a solid phase. A fresh sample was thus
subjected to the thermal protocol displayed in Fig. \ref{fig:Fig1}. It
has been reported that the formation of ordered solid phases in
glycerol is favored by a slow cooling to temperatures slightly above \textit{$T_{g}$} \cite{Mobius}, followed by two successive isothermal treatments
for triggering nucleation and crystal growth respectively. In order to demonstrate the formation of crystals seeds during the
thermal treatment at 190 K, selected measurements at this temperature and comparisons of the spectra at 230 K before and
after such nucleation are presented in Fig. \ref{fig:Fig3}(top) and Fig. 
\ref{fig:Fig3}(bottom) respectively. During nucleation, there is a slight
decrease of the permittivity. Whether this reduction is a consequence of
changes at the molecular level associated with crystal
nucleation or simply slow geometrical adjustments of the capacitor due
to mismatch of sample and spacer thermal-expansion coefficients is hard to
elucidate. Nevertheless, the bottom panel of Fig. \ref{fig:Fig3} proves the sample did not
undergo any kind of crystal growth at 190 K as indicated by the almost identical spectrum before and after such nucleation period.
\\

\begin{figure}
	\centering
	\includegraphics[trim = 10mm 65mm 10mm 70mm, clip, width=0.95\linewidth]{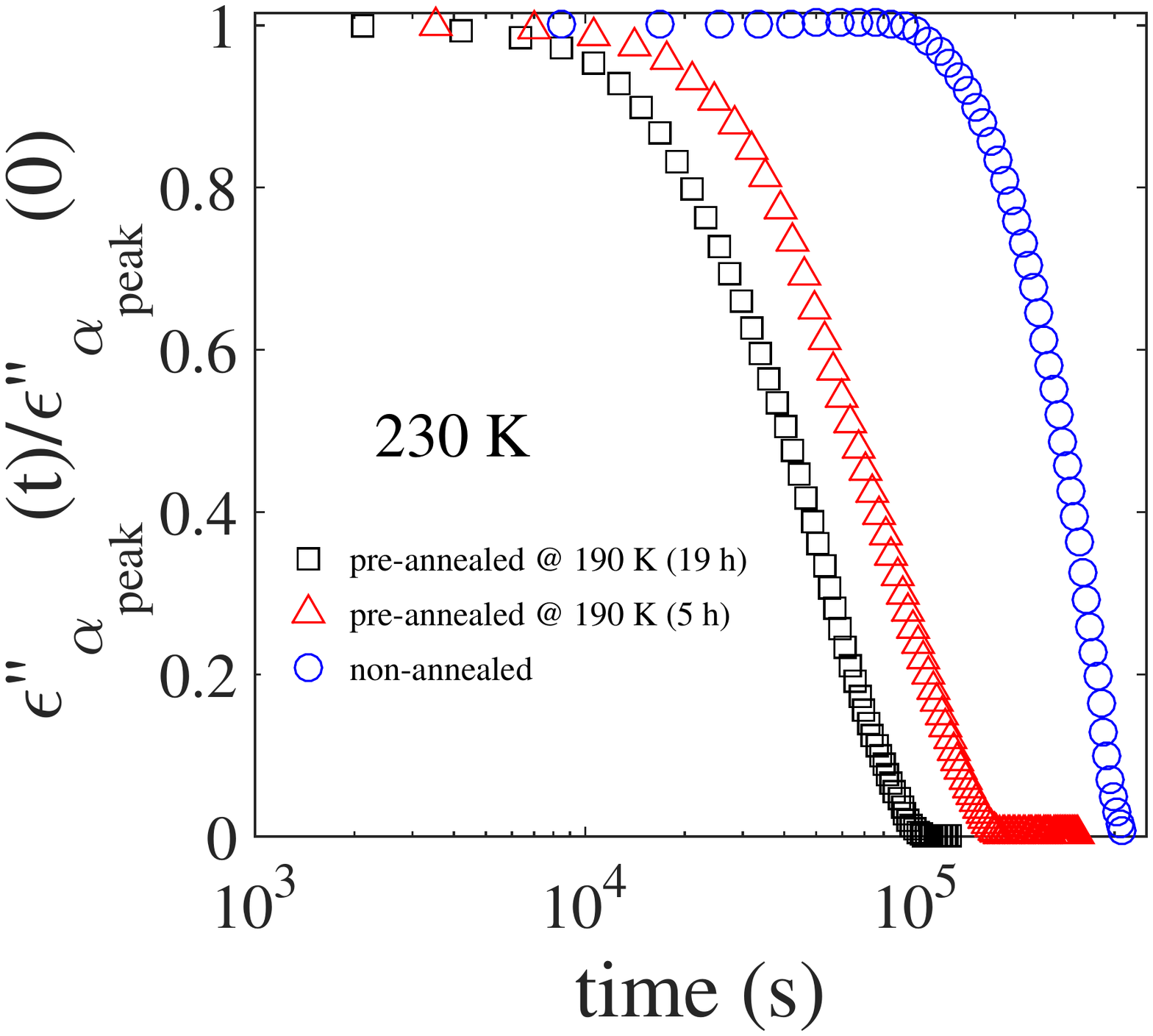}
	\caption{Time evolution of the maximum intensity of the loss peak upon crystallization at 230 K for samples without annealing (blue circles) or with annealing at 190 K during 5 h (red triangles) or for 19 h (black squares). Data are normalized to the initial value.}
	\label{fig:Fig5}
\end{figure}

In Fig. \ref{fig:Fig5} we show the
kinetics of crystallization at 230 K for different nucleation conditions. The non-annealed sample was directly cooled from room
temperature to 230 K at a cooling rate of 5 K/h. The data in
Fig. \ref{fig:Fig5} indicate that crystal growth becomes faster
with longer periods of annealing at 190 K. This finding implies the
formation of crystalline nuclei at 190 K which subsequently organize
into the crystal lattice at 230 K. We interpret this as an indication that there is in fact a structural difference between samples before and after annealing at 190 K. The fact that the two dielectric signals are identical tells us that the formation of nuclei is a very
local process that does not have any global effect on the liquid
dynamics.

\subsection{Liquid dynamics during crystal growth}

\begin{figure}
	\centering
	\includegraphics[trim = 45mm 75mm 25mm 72mm, clip, width=1.0\linewidth]{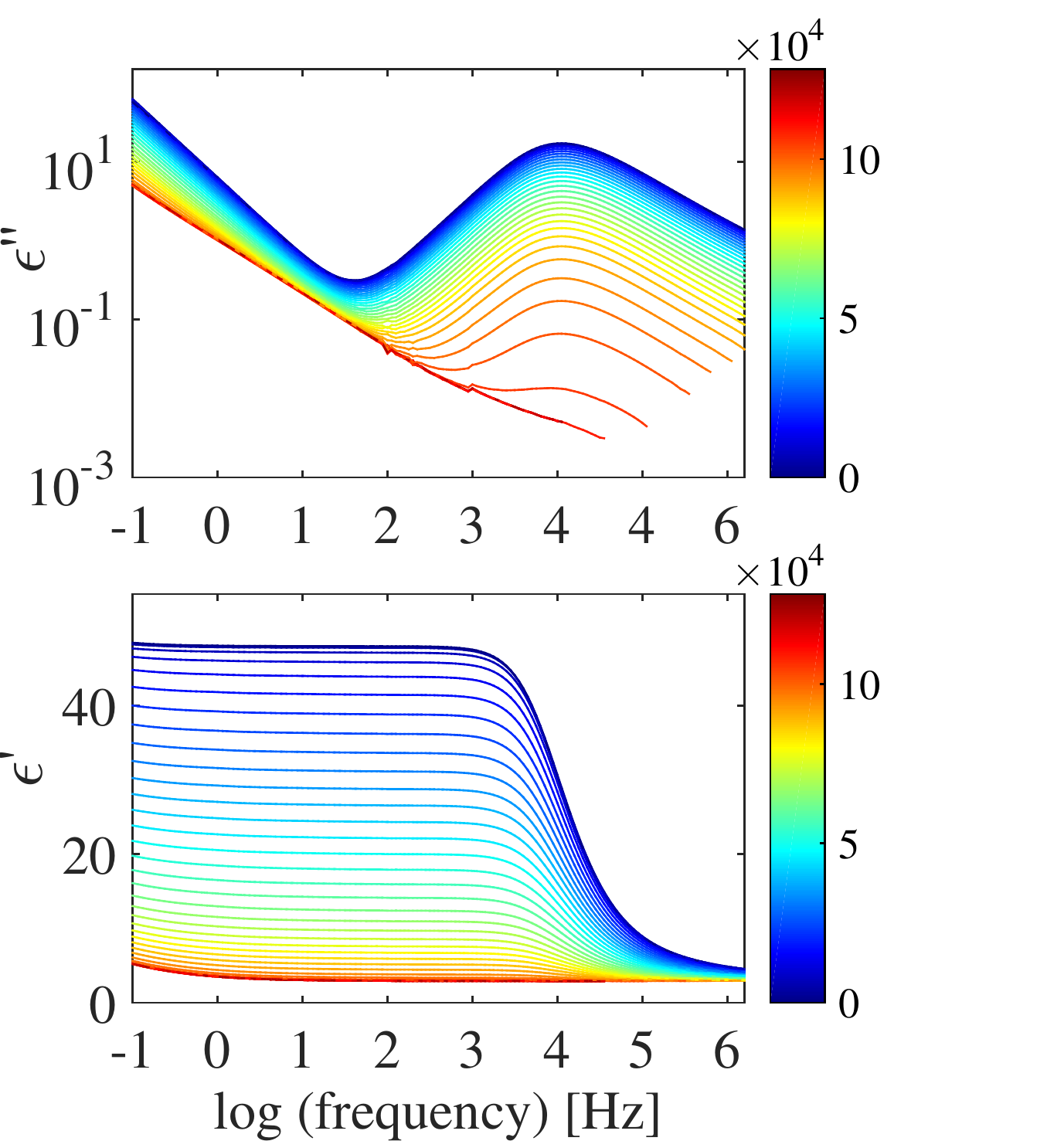}
	\caption{Complex dielectric permittivity of supercooled glycerol upon crystallization at 230 K. Imaginary (top) and real part of the permittivity (bottom) are represented as a function of frequency at different stages of crystallization. Colorbars indicate the crystallization time in seconds.}
	\label{fig:Fig6}
\end{figure}

With the purpose of studying the influence of crystallization on the
cooperative dynamics in glycerol, frequency sweeps from 100 mHz to 1
MHz (total acquisition time 11.7 min.) were collected continuously at
a constant temperature of 230 K. Selected snapshots of this real time
investigation are presented in Fig. \ref{fig:Fig6}. The dielectric
relaxation response remains unchanged during the early stages, but its
intensity starts to monotonically decrease after an induction period
and finally the $\alpha$ relaxation peak totally vanishes. According to the theory of dielectric relaxation and ignoring
possible variations of the dipole-dipole correlations, we may consider
the following proportionality

\begin{equation}
	\Delta\varepsilon\propto\frac{\mu^{2}N}{k_{b}T}
	\label{eq:Eq2} 
\end{equation}
where $\mu$ is the molecular dipole moment, \textit{$k_{b}$} is the
Boltzmann constant, \textit{T} is the temperature and \textit{N}
corresponds to the total number of reorientating dipoles in the system
\cite{Kremer}. Assuming that the transformation of a liquid substance
into a crystal carries with it the corresponding reduction of
\textit{N} fluctuating species as molecules abandon the disordered
phase and attach to the surface of growing crystals, where molecules
become translational and rotationally immobile, one accordingly expects
a depletion of $\Delta\varepsilon$ via Eq. \ref{eq:Eq2}. This effect
is clearly observed in both components of the complex permittivity as
displayed in Fig. \ref{fig:Fig6}. The low frequency dispersion
detected in $\varepsilon'' (\omega)$, which is assigned to pure ohmic
conduction, decreases approximately at the same rate as the $\alpha$
relaxation peak. A reduction of the dc conductivity upon
crystallization is in agreement with arguments that inversely
correlate the conduction of free charges with the viscosity of the
medium according to the Debye-Stokes-Einstein relation \cite{Roland}. It is important to remark that crystalline glycerol shows a shear viscosity larger than the one for the supercooled liquid \cite{Mobius}. Therefore we expect a larger viscosity for the composite material as crystallization proceeds. This would explain why the dc-conductivity decreases in spite of the $\alpha$ relaxation time remains unchanged. Besides, there is a significant modification of the charge-transport mechanism as revealed by the onset of a sublinear power law at low frequencies as crystallization proceeds. A detailed discussion of this crossover is beyond the scope of the present manuscript, but it deserves more attention in the future.
Apart from the total extinction of the
structural relaxation when the crystallization ended, it is important
to remark that the location of the peak remained unchanged during the
whole process, suggesting that the intrinsic nature of the structural
relaxation of glycerol remains unaltered in the course of
crystallization. A more detailed work on the
temperature dependence of the kinetics of crystallization has also
been carried out and it will be the subject of a future publication
\cite{preparation}. We have also considered possible Maxwell-Wagner-type
\cite{Mikkel} effects on the dielectric signal.

\begin{figure}
	\centering
	\includegraphics[trim = 25mm 45mm 20mm 52mm, clip, width=0.8\linewidth]{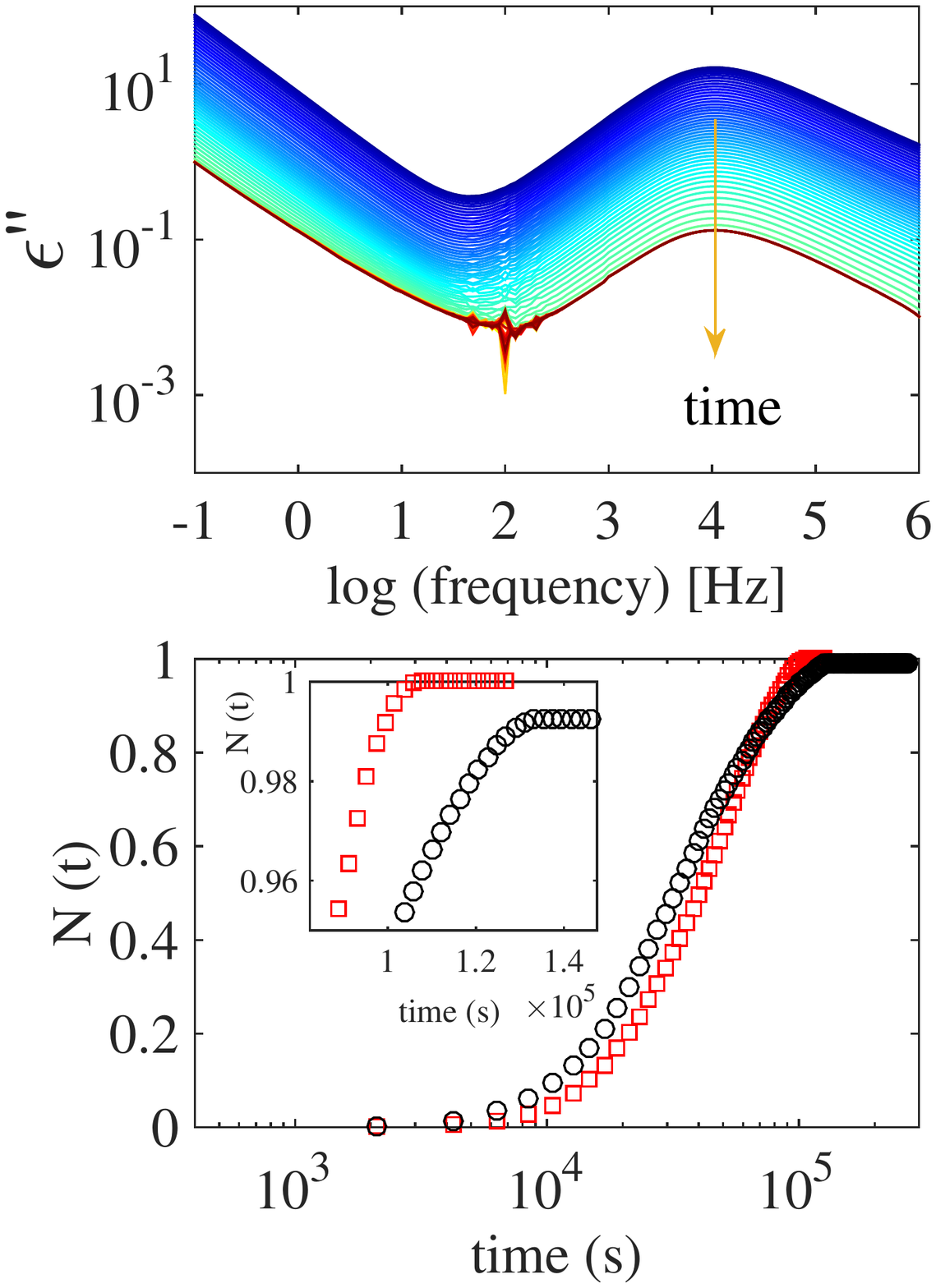}
	\caption{(Top) Dielectric loss in logarithmic scale of the $\alpha$ relaxation as a function of frequency at different crystallization times during isothermal annealing at 230 K. Snapshots are shown every 3.5x$10^3$ s during a total crystallization time of 2.7x$10^5$ s. (Bottom) Time dependence of the crystalline volume fraction at 230 K for the complete (red squares) and aborted (black circles) crystallization processes. The inset zooms the late stages to highlight the frustrated crystallization shown in the top panel.}
	\label{fig:Fig7}
\end{figure}

\begin{figure}
	\centering
	\includegraphics[trim = 50mm 75mm 25mm 72mm, clip, width=1.0\linewidth]{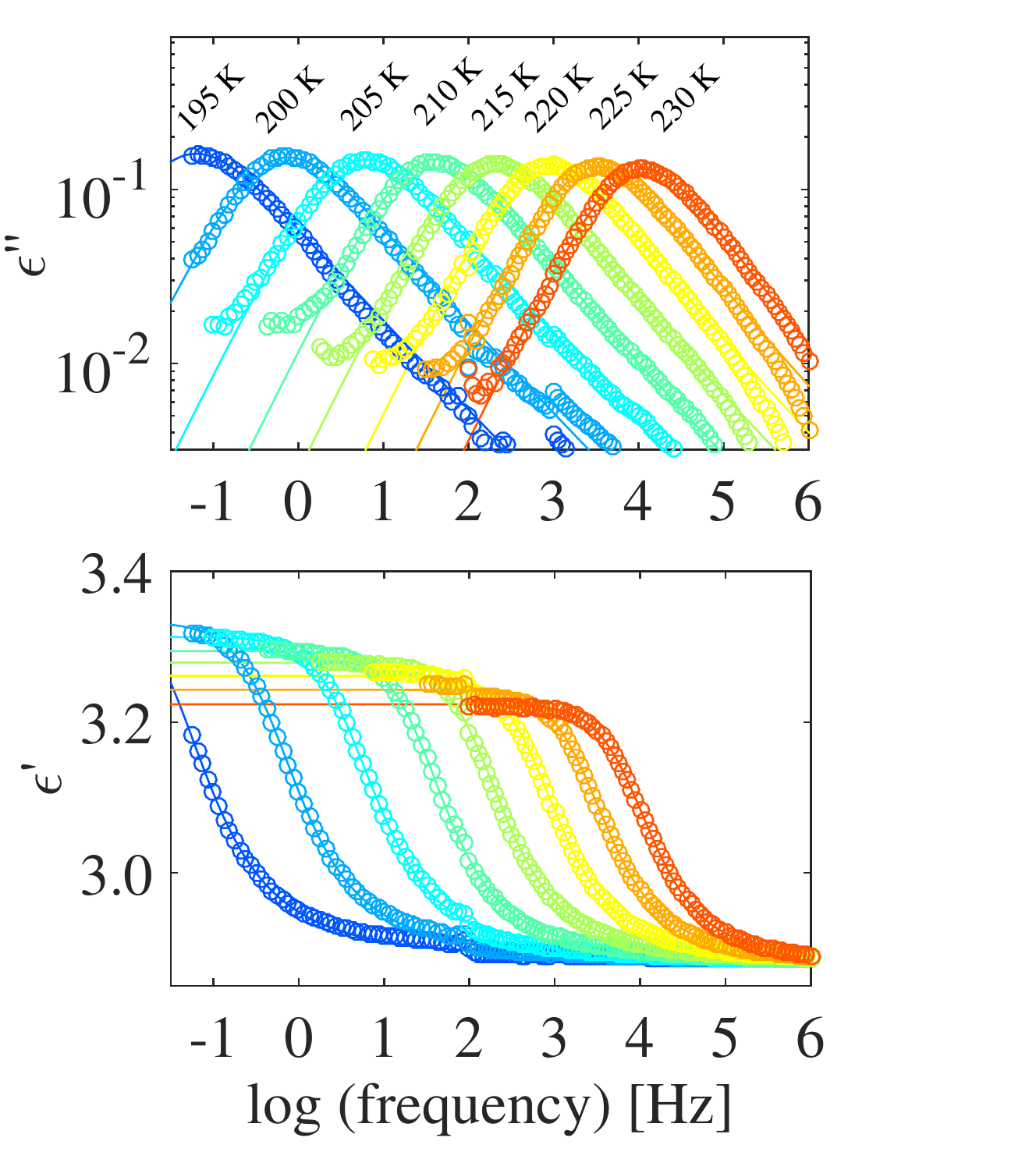}
	\caption{Dielectric spectra of residual liquid glycerol in a frustrated crystalline phase at the labelled temperatures. Continuous lines represent the best fit of the experimental data to the EB model.}
	\label{fig:Fig8}
\end{figure}

\subsection{Liquid dynamics after aborted crystallization}

We explored the crystallization of glycerol at different temperatures
several times. In one single case, the ordering process did
not proceed to the end, and a small fraction of liquid phase ($\sim 1 \%$) got
trapped between the crystalline domains. To illustrate this aborted
crystallization, Fig. \ref{fig:Fig7}(top) shows the evolution of the
$\alpha$ relaxation as a function of crystallization time. Contrary to
the data set shown in Fig. \ref{fig:Fig6}, a
residual and stable $\alpha$ peak is still detected once the
crystallization terminates. Such coexistence between crystallites and
disordered domains giving rise to an $\alpha$ peak is a well-known and
general phenomenon in the field of semi-crystalline polymers
\cite{Aurora}. Here, the stability of this residual relaxation was
confirmed over more than 24 hours without any indication of losing
intensity, broadening or shifting in frequency. 

The kinetics of crystallization was evaluated by calculating the
volume fraction of the new phase at different crystallization times $N(t)$
from the maximum intensity of the loss peak by means of
the following expression:

\begin{equation}
	N(t)=\frac{\varepsilon''_{\alpha_{peak}}(0)-\varepsilon''_{\alpha_{peak}}(t)}{\varepsilon''_{\alpha_{peak}}(0)-\varepsilon''_{\alpha_{peak}}(\infty)}
	\label{eq:Eq3}
\end{equation}
where $\varepsilon''_{\alpha_{peak}}(0)$ is the value of the dielectric loss at the frequency where the $\alpha$ peak is located ($log(f) = 4.05$ $Hz$) for the pure liquid, $\varepsilon''_{\alpha_{peak}}(t)$ takes the corresponding values at different crystallization times and $\varepsilon''_{\alpha_{peak}}(\infty)$ corresponds to the value of the dielectric loss at the same frequency for the pure crystal. For the specific case where the crystallization stopped and the exact value of $\varepsilon''_{\alpha_{peak}}(\infty)$ is unknown, we assume the same ratio between $\varepsilon''_{\alpha_{peak}}(0)$ and $\varepsilon''_{\alpha_{peak}}(\infty)$ obtained for the data at 230 K shown in Fig. \ref{fig:Fig6}.

Figure \ref{fig:Fig7} (bottom) shows the time evolution of $N(t)$ during isothermal treatment at 230 K for the complete (red squares) and aborted (black circles) phase transformations. In both cases, $N(t)$ follows a sigmoidal fashion in accordance with previous studies on crystalline growth \cite{Descamps}. In general terms, the kinetics of the transitions are quite similar, and one only detects significant differences close to the end where the maximum value of the volume fraction of the new phase does not reach 1 in one of the cases.\\

With the purpose of elucidating the dynamic properties of the liquid phase in
the frustrated crystal, the sample described in
Fig. \ref{fig:Fig7}(top) was subsequently cooled down to temperatures
just above \textit{$T_{g}$}, collecting isothermally every 5 K the
frequency dependence of the complex permittivity.

Full lines in Fig. \ref{fig:Fig8} correspond to the fit of the experimental data to the EB or $\alpha$-circuit model (Eq. \ref{eq:Eq1}). In both cases, the values of $\alpha$ were set to 0.5 for the whole temperature range. Since we were just interested in the $\alpha$ peak and due to the limited frequency range, the slight increase of dispersion at the high frequency flank of the $\alpha$ peak, the so-called excess wing, was not taken into consideration \cite{Lunkenheimer}.

\begin{figure}
	\centering
	\includegraphics[trim = 5mm 0mm 5mm 0mm, clip, width=0.9\linewidth]{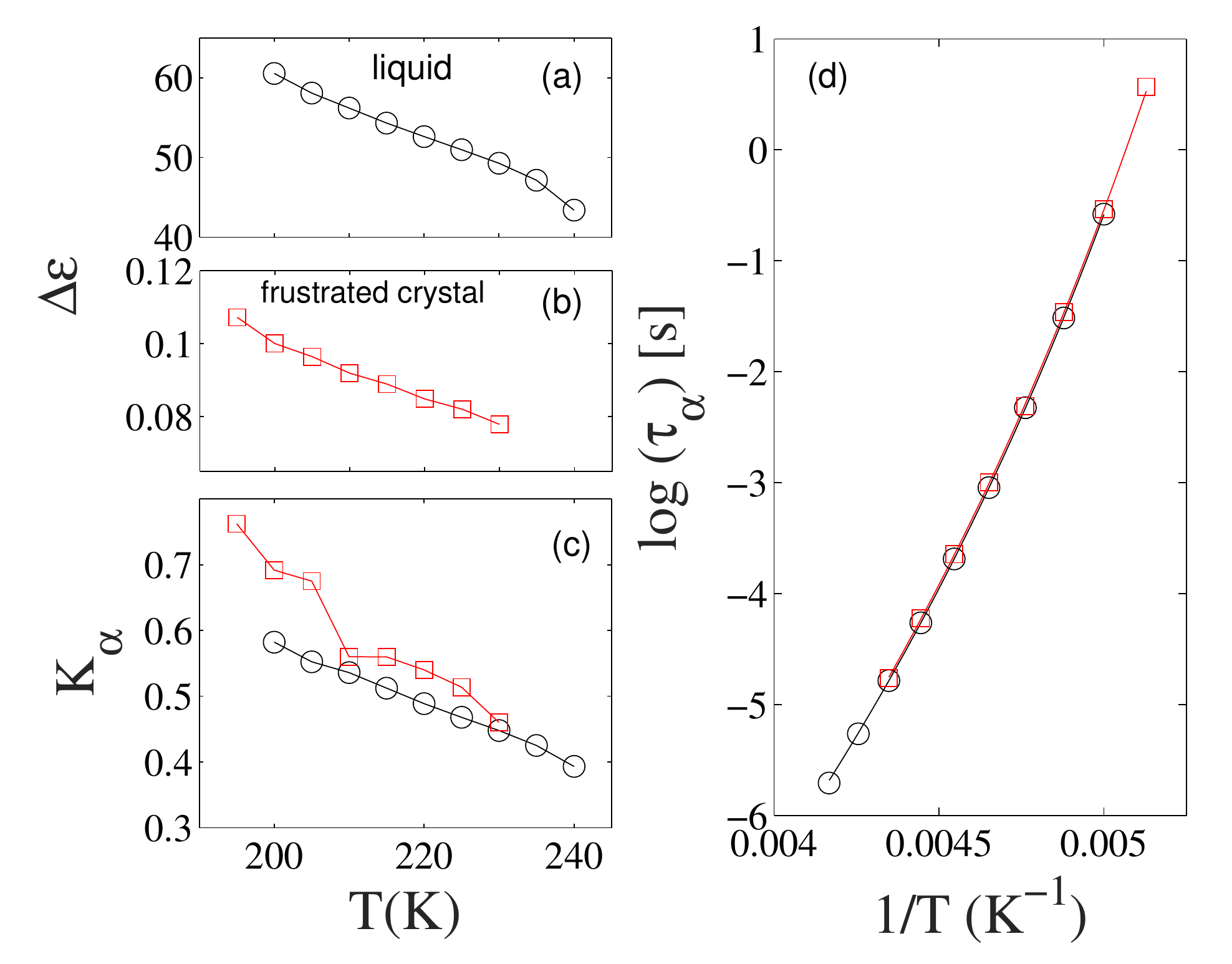}
	\caption{Temperature dependence of the fitting parameters of the EB model (Eq. \ref{eq:Eq1}) for the pure liquid (black circles) and frustrated crystalline glycerol (red squares). Panels $\textit{a}$ and $\textit{b}$ display the dielectric strength $\Delta\varepsilon$ and panel \textit{c} the broadening parameter $k_{\alpha}$. In panel $\textit{d}$ we show the characteristic relaxation times of the $\alpha$ process with reciprocal temperature. Solid lines in $\textit{d}$ are fits to the VFT equation. Lines in $\textit{a}$, $\textit{b}$ and $\textit{c}$ are guides to the eye.}
	\label{fig:Fig9}
\end{figure}

A close comparison of the fitting parameters of Eq. \ref{eq:Eq1} for
the liquid and aborted crystalline samples is shown in
Fig. \ref{fig:Fig9}. Regarding the broadening parameter $k_{\alpha}$,
Fig. \ref{fig:Fig9}(c) presents a similar behaviour for both kind of
liquids. The values of $k_{\alpha}$ increase when temperature
decreases which indicates a broader width of the distribution of
relaxation times when approaching \textit{$T_{g}$}
\cite{Lunkenheimer}. The values of $k_{\alpha}$ account for the
departure of the relaxation width from the simplest Debye model in
such a way that large values of $k_{\alpha}$ will lead to a less
Debye-like relaxation \cite{Debye,Saglanmak}. For the frustrated
crystal one may observe slightly higher values. It is tempting to
correlate this increase of $k_{\alpha}$ with a more heterogeneous
landscape. However, the difference is not very prominent and we have
to take into consideration that the amplitude of the relaxation for
the frustrated crystal is very low, which means that the curve shape is
also determined with less precision. The evolution of $\Delta\epsilon$ with temperature shows a similar trend for the liquid and frustrated crystal. In both cases, $\Delta\epsilon$ decreases with temperature in qualitative agreement with previous reports on supercooled glycerol \cite{Lunkenheimer}.

In Fig. \ref{fig:Fig9}(d) we have plotted the characteristic
relaxation times of the $\alpha$ process in the Arrhenius
representation. As expected, the temperature dependence of the
$\alpha$ relaxation times deviates from linearity as temperature goes
down. For both types of liquid glycerol, the values of $\tau_{\alpha}$
can be described by the empirical Vogel-Fulcher-Tammann (VFT)
relation:

\begin{equation}
	\tau_{\alpha}=\tau_{0}\exp\left(\frac{DT_{0}}{T-T_{0}}\right)     
	\label{eq:Eq4}
\end{equation}
where $\tau_{0}$ is a pre-exponential factor with phonon-like time
scales, \textit{D} is a strength parameter related to the dynamic
fragility through the following expression \textit{m=16+590/D} and
\textit{$T_{0}$} is the Vogel temperature ($T_{0}$ $<$ $T_{g}$)
\cite{Bohmer}. Leaving
aside the physical meaning of the parameters in Eq. \ref{eq:Eq4} (see
eg. \cite{Tina} for a discussion), the
usefulness of the VFT law for interpolating experimental points and
quantifying the steepness of the relaxation plot is widely accepted
\cite{Ediger,Debenedetti,Berthier,Tina}. Solid lines in
Fig. \ref{fig:Fig9}(d) correspond to the fits of the data to the VFT
equation for the pure liquid and frustrated crystalline samples. For
consistency, the relaxation times for the two kinds of liquids were
fitted in the same temperature range (200-230 K). Both the data and
the VFT lines practically overlap, which indicates that
the nature of the $\alpha$ relaxation dynamics is not significantly
altered by the presence of crystals. Similar values of the VFT parameters to those collected
here in Table \ref{tab:Tab1} have been reported in the literature
\cite{Lunkenheimer,Sudo}.

\begin{table}
	\caption{\label{tab:journals}VFT parameters of the $\alpha$ process for the pure liquid and for the remaining liquid fraction embedded in the frustrated crystal.}
	\begin{ruledtabular}
		\begin{tabular}{lll}
			\textbf{VFT Parameter} & \textbf{Pure Liquid}&\textbf{Frustrated Crystal} \\
			log $\tau_{0}$ (s) &\texttt{-14.8}&-14.9\\
			D  &\texttt{18.3}&\texttt{18.6}\\
			T$_{0}$ (K) &\texttt{128}&\texttt{128}\\
		\end{tabular}
	\end{ruledtabular}
	\label{tab:Tab1}
\end{table}

\section{Discussion}

We have explored the formation of solid structures in real time
in the well-known glass former  glycerol. The solidification was
induced by a two-step annealing procedure as suggested before
\cite{Mobius,Yuan}. Our dielectric data reveal that liquid glycerol,
through a nucleation step at 190 K and subsequent crystal growth at
230 K, transforms into a new phase. In most cases (not all of them
shown here \cite{preparation}), we have found a total disappearance of
the $\alpha$ relaxation, a fact which suggests that glycerol
transforms into the standard orthorhombic crystalline phase
\cite{Bermejo}. The fact that in one of the samples studied here
(Fig. \ref{fig:Fig7}) the end of the crystallization was suddenly
aborted could explain prior results that speculated the
formation of a glacial phase in which metastable nano-crystals were
embedded in a liquid matrix \cite{Mobius,Zondervan,Yuan}. Our data
just show an unfinished ordering process that resulted in a small
fraction of sample remaining in the liquid state. Unlike in polymeric
materials where the formation of semi-crystalline structures is a
common feature, few examples can be found in the literature of a
low molecular weight liquid that does not crystallize into a 100 $\%$
crystalline system \cite{Descamps}.

We have presented evidence that the nature of the $\alpha$ relaxation dynamics
does not undergo important modifications during crystallization. We
have extended our study along the whole crystallization process, from
a well controlled nucleation step to the crystal
growth. Interestingly, the structural relaxation remained unaffected,
first by the density fluctuations that trigger the formation of
nuclei, and second during the thickening of the lattice structure.

Taking advantage to the aborted crystallization that took place in one
of the studied samples, we could explore the temperature evolution of
the $\alpha$ relaxation associated with a small fraction of liquid phase
surrounded by the crystalline network. Our results indicate that the
dynamics remained virtually identical to the bulk liquid in terms of
location of the $\alpha$ peak, spectral shape and activation energy.
A relaxation process in crystalline glycerol was reported by Ryabov et
al. \cite{Ryabov}, where a sample of dry glycerol was crystallized on
heating, giving as a result a new dielectric mode five orders of
magnitude slower than the $\alpha$ process for the pure liquid. This
new mode exhibited an Arrhenius behaviour, similar to other dielectric
relaxation processes detected in molecular crystals,
\cite{Sanzacetone,Winkel} and it was assigned by the authors to either
defects in the crystalline lattice or to the presence of disordered
domains located at the interface between crystalline grains. In
contrast to the observation by Ryabov et al. \cite{Ryabov}, here we
detect a relaxation with the same features as the one showed by the
pure liquid, in such a way that we may confirm that it does not
correspond to the relaxation of crystalline glycerol itself, but to a
kind of metastable semi-crystalline system. We do not see any evidence
of the slow mode reported by by Ryabov \emph{et al}. \cite{Ryabov}.

Several factors have been proposed to affect the mobility of the
liquid at the proximities of the crystal/liquid interfaces during
crystallization, including the existence of pre-ordered configurations
and the different ability of the molecules to be accommodated on the
surface of the growing lattice \cite{Descamps}. In the latter case, it is expected
that not all molecules near the crystal/liquid interface show the same
geometrical preferences to attach to the crystal. For both cases, the
activation energy that governs diffusion over short distances may
increase as compared to the pure liquid \cite{Descamps}. Here we
report that glycerol, at least for the time scale detected by dielectric spectroscopy, does not present
variations in its mobility upon crystallization.
\\
The molecules in
crystalline glycerol are in an extended conformation in which the two
dihedral angles defined by the intersection between two specified
planes within the molecule are situated in a range of 120$^{\circ}$
and 240$^{\circ}$ \cite{Bermejo}. On the contrary, Towey et
al. \cite{Towey} have recently reported that in the liquid state the
most probable conformer shows a less extended structure in which the
second dihedral angle presents values in the range
240-360$^{\circ}$. Therefore, the transference of units from the
liquid on to the surface of the growing crystal should imply, at least
for an important fraction of molecules, the necessity of
conformational transitions to the extended or coplanar
configuration. However, according to our data, this does not cause any
global effect on the dynamics. The lack of dynamical signature of the
crystallization process indicates that the changes in the hydrogen bonding
network as well as the conformational changes taken place during the
crystallization are very local. Thus, very few molecules at the phase
boundary are changing structure while the majority of the molecules
are either in the crystal or in the ordinary bulk liquid.

\section{Conclusions}
We present, to the best of our knowledge, the first study in real time
of the isothermal crystallization of glycerol by using dielectric
spectroscopy. We demonstrate that liquid glycerol transforms into a
crystalline material by following a two-step nucleation and crystal
growth protocol. Our results do not support the formation of a
distinct solid phase from the standard crystalline lattice. We have
also shown that in one of the samples studied, the starting liquid
transformed into an incomplete crystalline phase with a residual
liquid fraction. For a period of more than one day we proved the
metastability of this frustrated crystal. The dynamics of this
remaining liquid phase were studied as a function of temperature
showing features very close to the pure liquid material. In addition
to that, we have shown that during the ordering process the
$\alpha$ relaxation dynamics remains unaltered, revealing that neither
the nucleation period nor the crystal growth affected the nature of
the cooperative motions in supercooled glycerol.

\begin{acknowledgments}
	This work has been funded by the Danish Council for
	Independent Research (Sapere Aude: Starting Grant). Technical support from the workshop at IMFUFA (Department of Science and Environment, Roskilde University) is acknowledged. We also thank Michael Greenfield for technical support.
\end{acknowledgments}

\bibliography{glycerol}

\begin{thebibliography}{47}
\bibitem [{alp()}]{alpha}%
  \BibitemOpen
  \href@noop {} {\bibinfo  {journal} {It is well documented that the dielectric
  loss of the $\alpha$ relaxation of viscous molecular liquids at the high
  frequency limit shows a power law dependence of $\omega^{-\alpha}$, where
  $\alpha$ normally takes values of 0.5 when temperature approaches the glass
  transition}\ }\BibitemShut {NoStop}%
\end{thebibliography}

\end{document}